\documentclass[12pt,A4,]{article}
\usepackage[margin=1in,footskip=0.25in]{geometry}
\usepackage{latexsym}
\usepackage{amsmath,mathrsfs,amsfonts,amssymb,bm,mathtools}
\usepackage{graphicx, algorithm, algorithmic}

\title{\bf\Large Interval-valued Data Prediction via Regularized Artificial Neural Network}
\author{Zebin Yang$^1$, Dennis K.J. Lin$^2$ and Aijun Zhang$^1$\\
{\normalsize  $^1$Department of Statistics and Actuarial Science, The University of Hong Kong}\\
{\normalsize Pokfulam Road, Hong Kong}\\
{\normalsize $^2$Department of Statistics, Pennsylvania State University}\\
{\normalsize University Park, PA 16802, USA}}
\date{} 

\begin{document}
\maketitle

\begin{abstract}
A regularized artificial neural network (RANN) is proposed for interval-valued data prediction. The ANN model is selected due to its powerful capability in fitting linear and nonlinear functions. To meet mathematical coherence requirement for an interval (i.e., the predicted lower bounds should not cross over their upper bounds), a soft non-crossing regularizer is introduced to the interval-valued ANN model. We conduct extensive experiments based on both simulation datasets and real-life datasets, and compare the proposed RANN method with multiple traditional models, including the linear constrained center and range method (CCRM), the least absolute shrinkage and selection operator-based interval-valued regression method (Lasso-IR), the nonlinear interval kernel regression
(IKR), the interval multi-layer perceptron (iMLP) and the multi-output support vector regression (MSVR). Experimental results show that the proposed RANN model is an effective tool for interval-valued prediction tasks with high prediction accuracy.

\end{abstract}

\vskip 6.5pt

\section{Introduction}
Interval-valued data is almost everywhere in our daily life; for example,  climate records, stock prices and aggregation statistics from large datasets. Unlike traditional point values, the interval-valued data could naturally provide extra information for more precise decision making. However, the interval-valued prediction problem does not receive as much attention as the point prediction, despite its importance in real life. In this paper, we look into this interesting topic and propose an appealing interval-valued prediction model.

The center method (CM) proposed by Billard and Diday (2000) is the first linear regression model for interval-valued prediction. It assumes the lower
and upper bound of the interval share the same linear relationship among target variables $Y=\left\langle {Y^{L} ,Y^{U} } \right\rangle $
and explanatory variables ${\rm {\bf X}}=\left\langle {{\rm {\bf X}}^{L}, {\rm {\bf X}}^{U} } \right\rangle $. This relationship can be estimated by
fitting a linear regression using the corresponding centers $Y^c =\frac{1}{2}(Y^{L} +Y^{U} )$ and ${\rm {\bf X}}^{c} =\frac{1}{2}({\rm {\bf
X}}^{L} +{\rm {\bf X}}^{U} )$. However, the CM method is limited due to its strict assumption, and it may be not an appropriate choice for real-world datasets. In addition, neither the CM method can ensure the mathematical coherence of predicted intervals, i.e., the predicted lower bounds should be smaller than the corresponding upper bounds.

Many studies have been conducted to improve the CM method. For example, Billard and Diday (2002) proposed the MinMax method, which fits two linear
regression models for the lower and upper bounds respectively. By transforming the original interval-valued prediction task into two
independent prediction tasks, i.e., center prediction and range prediction, Lima Neto and De Carvalho (2008) proposed the center and range method (CRM),
which could utilize more information compared with the CM method. To ensure the mathematical coherence of predicted intervals, Lima Neto and De
Carvalho (2010) further proposed the constrained center and range method (CCRM) by adding a non-negative constraint on the coefficients of range regression model.
Giordani (2015) proposed a least absolute shrinkage and selection operator-based interval-valued regression (Lasso-IR) method, which
is established on the constraint that guarantees the positiveness of the estimated range. Similarly, Hao and Guo (2017) proposed a constrained center
and range joint model (CCRJM), which also considered the positiveness of the estimated range.

The linear model is too restrictive for the complicated real-world interval-valued datasets. For example, the stock market and climate systems
are both the places where mass interval-valued data exist. These data are widely known for their nonlinearity and uncertainty. For nonlinear interval
data, the nonparametric method is a popular choice. To name some recent studies, Fagundes et al.~(2014) proposed the interval kernel regression
(IKR) method, by reformulating the CRM method in kernel smoothing settings. Jeon et al.~(2015) proposed a joint empirical distribution estimator of intervals via the Gaussian kernel and used it for interval-valued prediction. Lim (2016) considered using a nonparametric additive model for interval-valued
data, which is more suitable for handling nonlinear patterns.

Besides nonparametric methods, the machine learning approach can be applied to interval-valued data prediction. Due to the complex model structure and good performance in practical applications, machine learning algorithms have become increasingly popular and many researchers have introduced them for interval-value prediction purposes. For instance, San Roque et al.~(2007) proposed an interval multi-layer perceptron (iMLP) model, considering the center and range regression using the neural network framework. Xiong et al.~(2014a) introduced a multi-output support vector regression (MSVR) method in interval-valued time series forecasting problem. Some relevant work can also be referred to the artificial neural networks (ANN) for the center and range regression (Maia et al., 2008), ANN for lower and upper bounds regression (Maia and de Carvalho, 2011), support vector regression (SVR; Xiong et al., 2014b). Despite the high accuracy of these machine learning-based models, the mathematical incoherence of the predicted interval did not get as much attention. For example, the ANN, SVR and MSVR models cannot prevent interval crossing problem. To avoid the interval crossing, the iMLP model uses absolute-valued weights for range regression. However, this would make the iMLP model much more difficult in model training. More detailed discussion of the iMLP model can be found in Section 2.

In this paper, we develop a machine learning-based prediction model for interval-valued data, the regularized artificial neural network (RANN). In RANN, a non-crossing regularizer for preventing interval incoherence is incorporated with the ANN model to tackle the interval-valued data prediction task. This model is characterized by its ability in handling nonlinear and incoherence for interval-valued data. First, unlike the iMLP method reviewed above, the proposed RANN model treats all independent variables' lower and upper bounds as input to predict the target interval. Under this setting, these two bounds would share the same hidden layers (commonality) but differ in the output layer (specialty). Thus, this model is thought to be able to capture both individual behavior and cross-correlation between the upper and lower bounds. Second, using a soft non-crossing regularizer, both goals of prediction accuracy and mathematical coherence could be achieved simultaneously, without oversacrificing model performance. Moreover, by tuning the regularization parameter, the RANN model can be flexibly adjusted according to different datasets.

The rest of the paper is organized as follows. In Section 2, we review a selective set of classical interval-valued  prediction methods from linear, nonparametric and machine learning-based perspectives. The proposed RANN model is presented in Section 3 with discussions on model formulation and model training. The experimental studies with simulation data and real-life data are presented in Section 4. Section 5 concludes the paper with remarks.

\section{Review of Existing Methods}
\subsection{Constraint Center and Range Method}
The center and range method (CRM) is proposed by Lima Neto and De Carvalho
(2008), and it has become one of the most important models for analyzing
interval-valued data. This method follows the idea of the center method (CM;  Billard and Diday, 2000) and decomposes the interval-valued data
prediction task into two independent subtasks; i.e., the center linear regression and range linear regression. Utilizing more information, the CRM
method is generally thought to be more accurate than the CM method.

Define an interval-valued data $({\rm {\bf X}}_k ,Y_k )$ for $k=1,2,...,N$
with ${\rm {\bf X}}_k =\left\langle {{\rm {\bf X}}_k^L ,{\rm {\bf X}}_k^U }
\right\rangle $ and $Y_k =\left\langle {Y_k^L ,Y_k^U } \right\rangle $,
where ${\rm {\bf X}}_k^L ,{\rm {\bf X}}_k^U $ are $p$-dimensional
independent variables denoting the lower and upper bounds, respectively. In
CRM, the interval-valued data is first transformed into centers and
half-ranges:
\begin{equation} \label{eq1}
\def\arraystretch{1.3}
\begin{array}{l l}
{\rm {\bf X}}_k^c =\frac{1}{2}({\rm {\bf X}}_k^L +{\rm {\bf X}}_k^U), & {\rm {\bf X}}_k^r =\frac{1}{2}({\rm {\bf X}}_k^L +{\rm {\bf X}}_k^U ), \\
 Y_k^c =\frac{1}{2}(Y_k^L +Y_k^U), & Y_k^r =\frac{1}{2}(Y_k^L +Y_k^U), \\
 \end{array}
\end{equation}
where ${\rm {\bf X}}_k^c =(X_{k1}^c ,X_{k2}^c ,...,X_{kp}^c), Y_k^c$
denote the centers, and ${\rm {\bf X}}_k^r =(X_{k1}^r
,X_{k2}^r ,...,X_{kp}^r), Y_k^c$ are the corresponding half-ranges. The CRM
method fits two linear regression models,  
\begin{equation}\label{eq2}
\def\arraystretch{1.3}
\begin{array}{l}
 Y_k^c =\beta _0^c +\beta _1^c X_{k1}^c +...+\beta _p^c X_{kp}^c
+\varepsilon _k^c , \\
 Y_k^r =\beta _0^r +\beta _1^r X_{k1}^r +...+\beta _p^r X_{kp}^r
+\varepsilon _k^r . \\
 \end{array}
\end{equation}
Using matrix notations, the ordinary least square (OLS) method can be used to
solve this problem (assuming full rank):
\begin{equation}
\label{eq3}
{\rm {\bf \hat {\beta }}}^c =(({\rm {\bf X}}^c)^T{\rm {\bf X}}^c)^{-1}({\rm
{\bf X}}^c)^T{\rm {\bf Y}}^c\mbox{ and }{\rm {\bf \hat {\beta }}}^r =(({\rm
{\bf X}}^r)^T{\rm {\bf X}}^r)^{-1}({\rm {\bf X}}^r)^T{\rm {\bf Y}}^r.
\end{equation}
To tackle the interval crossing problem, Lima Neto and De Carvalho (2010)
proposed a constraint center and range method (CCRM), where a non-negative
constraint$\mbox{ }{\rm {\bf \beta }}^r \ge 0$ is added in range linear
regression. Thus, the estimated range $\hat {Y}^r $ would always be
positive, which   ensures the mathematical coherence.

This CCRM model is effective in many scenarios. However for some complicated problems where nonlinear patterns exist, the linear regression models do
not perform well. To meet the challenge of nonlinear interval-valued data prediction problems,  researchers have proposed to use nonparametric regression  and machine learning-based methods.

\subsection{Interval Kernel Regression}
Kernel method is a popular nonparametric modeling tool for datasets without explicit distribution information. Fagundes et al.~(2014) introduced the interval kernel regression (IKR) method for interval-valued data predictio, and the most representative version is the IKRCR method based on center and range information.

Similar to the CRM method, the IKRCR method models the centers and ranges seperately. For the $k$-th sample, the Gaussian kernel functions are computed by
\begin{equation}\label{eq4}
K({\rm {\bf X}}^c ,{\rm {\bf X}}_k^c )=\left( {\frac{1}{\sqrt {2\pi } h}}
\right)^pe^{-\frac{\left\| {{\rm {\bf X}}^c -{\rm {\bf X}}_k^c }
\right\|^2}{2h^2}}\mbox{ and }K({\rm {\bf X}}^r ,{\rm {\bf X}}_k^r )=\left(
{\frac{1}{\sqrt {2\pi } h}} \right)^pe^{-\frac{\left\| {{\rm {\bf X}}^r
-{\rm {\bf X}}_k^r } \right\|^2}{2h^2}},
\end{equation}
where $h$ is a pre-specified bandwidth parameter.  Then, the center and range predictions can be obtained by  
\begin{equation}
\label{eq5}
\hat {Y}^c  =\sum\limits_{k=1}^N {w_k^c Y_k^c } \mbox{ and }\hat {Y}^r
=\sum\limits_{k=1}^N {w_k^r Y_k^r } ,
\end{equation}
with the weights $w_k^c $ and $w_k^r $ determined by
\begin{equation}
\label{eq6}
w_k^c =\frac{K({\rm {\bf X}}^c ,{\rm {\bf X}}_k^c )}{\sum\nolimits_{k=1}^N
{K({\rm {\bf X}}^c ,{\rm {\bf X}}_k^c )} }\mbox{ and }w_k^r =\frac{K({\rm
{\bf X}}^r ,{\rm {\bf X}}_k^r )}{\sum\nolimits_{k=1}^N {K({\rm {\bf X}}^r
,{\rm {\bf X}}_k^r )} }.
\end{equation}

\subsection{Interval Multi-layer Perceptron}
San Roque et al.~(2007) proposed the interval multi-layer perceptrons (iMLP) model for interval-valued data. Like the CRM method, the iMLP model tries to
solve the interval-valued prediction problem by fitting a center and range
regression using artificial neural network (ANN) architecture. There are two
main differences between iMLP and standard ANN. First, the input and output of each neuron in iMLP are center-range paired values,  so the center
and range units share the same connecting weights. Second, similar to the
idea in the CCRM method, the iMLP model uses absolute-valued weights  in range prediction, in order to guarantee the positiveness of the predicted ranges. An iMLP model with $p$ input neurons, $J$ hidden neurons,
and one output neuron can be represented as follows:
\begin{equation} \label{eq7}
\def\arraystretch{1.5}
\begin{array}{l l}
h_j^c ={\rm {\bf w}}_j^{(h)} {\rm {\bf X}}_i^c +b_j^{(h)}, & h_j^r =\left|
{{\rm {\bf w}}_j^{(h)} } \right|{\rm {\bf X}}_i^r ,\\
H_j^c =\frac{1}{2}\left[ {\tanh (h_j^c +h_j^r )+\tanh (h_j^c -h_j^r )} \right], & H_j^r =\frac{1}{2}\left[ {\tanh (h_j^c +h_j^r )-\tanh (h_j^c -h_j^r
)} \right],\\
\hat {Y}^c=\sum\limits_{j=1}^J {w_j^{(o)} } H_j^c +b^{(o)} , & \hat {Y}^r=\sum\limits_{j=1}^J {\left| {w_j^{(o)} } \right|H_j^r } ,
\end{array}
\end{equation}
where $\left\langle {h_j^c ,h_j^r } \right\rangle $ and $\left\langle {H_j^c, H_j^r } \right\rangle $ are the $j$-th hidden neuron's input and output,
respectively. The weights and bias connecting the input layer and the $j$-th
hidden neuron are denoted as ${\rm {\bf w}}_j^{(h)} $ and $b_j^{(h)} $, and
the output layer weight and bias are $w_j^{(o)} $and $b^{(o)} $.

Although the iMLP model could guarantee the mathematical coherence of predicted intervals,  these absolute operators would make the network hard to train and even not fall into local minima. Besides in iMLP, the center prediction and the range prediction share all the connecting weights.  This is equivalent to enforce that the lower and upper bounds follow identical data generation rules, which is however inappropriate for most real-life datasets. 

\section{The Proposed Method}
With the recent advances in artificial intelligence and big data, the machine learning methods, especially the neural network models, are receiving more and more attention, as compared with statistical regression models. One reason is their flexible model structure and powerful nonlinear representation. Several machine learning-based models have been successfully employed to solve interval-valued data prediction tasks, including the ANN, MSVR and iMLP models that we have reviewed in previous sections.

However, these existing machine learning models fail to provide a good balance for the prediction accuracy and interval crossing problem, where  the interval crossing occurs when the predicted lower bound greater than the predicted upper bound. This is
a violation of the basic interval property. Both linear and nonlinear models may have this problem. As discussed in the previous section, several linear models have been proposed to prevent from such problem, including the CCRM method (Lima Neto and De Carvalho, 2010), the Lasso-IR method (Giordani, 2015) and the CCRJM method (Hao and Guo, 2017). For the machine learning-based models, only the iMLP model made an attempt to deal with this crossing problem. We find that these models are all based on seperate modeling of centers and ranges, with additional inequality constraints to ensure the positiveness of the predicted ranges. Although the interval crossing problem could be mitigated, these inequality constraints may sometimes bring severe drawback to the prediction accuracy.

\begin{figure}[htbp]
\centerline{\includegraphics[width=4.5in,height=2.7in]{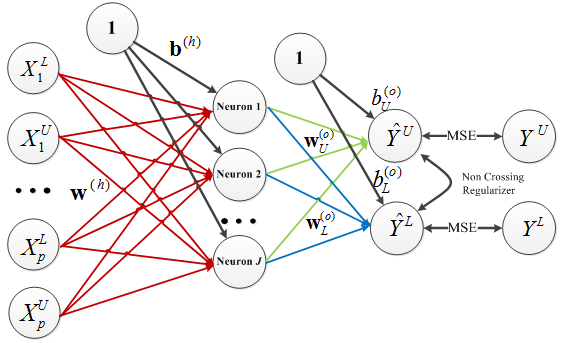}}
\caption{Architecture of the Proposed RANN Model}\label{fig1}
\end{figure}

In what follows, we propose a regularized artificial neural network (RANN) by introducing a soft non-crossing regularizer for interval-valued prediction, with network architecture shown in Figure~\ref{fig1}, 
which will be shown more flexible and effective in modeling nonlinear relationships of interval-valued data. 

\subsection{Model Formulation}
The proposed RANN model is based on a three-layer ANN model in which both the inputs and outputs contain the lower and upper bound values of the intervals. As shown in Figure~\ref{fig1}, the interval-valued data $({\rm {\bf X}},Y)$ are composed of $p$ independent variables and one target
variable for each of the lower and upper bound, as $\{X_1^L ,X_1^U
,...,X_p^L ,X_p^U \}$ and $\{Y^L,Y^U\}$. For convenience, we also use the
notation $X_i, i=1,2,...,2p$ for independent variables. The corresponding
three-layer ANN model has a structure of $2p$ input neurons, $J$ hidden
neurons, and 2 output neurons. The outputs of this model can be written as:
\begin{equation}\label{eq10}
\hat {Y}^L=f\left(\sum\limits_{j=1}^J {z_j w_{j,L}^{(o)} } +b_L^{(o)}\right)
\mbox{ and }
\hat {Y}^U=f\left(\sum\limits_{j=1}^J {z_j w_{j,U}^{(o)} } +b_U^{(o)}\right),
\end{equation}
where ${\rm {\bf w}}_L^{(o)} =\{w_{1,L}^{(o)} ,w_{2,L}^{(o)}
,...,w_{J,L}^{(o)} \}$ is the weight vector between the lower bound output
neuron and the hidden layer, ${\rm {\bf w}}_U^{(o)} =\{w_{1,U}^{(o)}
,w_{2,U}^{(o)} ,...,w_{J,U}^{(o)} \}_{ }$ denotes the corresponding weight
vector for upper bound output, and $b_L^{(o)} $, $b_U^{(o)} $ are their bias
terms. The activation function of the output layer is denoted as $f$. The
symbol $z_j $ represents the output of the $j$-th hidden layer neurons, which
has the following representation:
\begin{equation}\label{eq11}
z_j =g\left(\sum\limits_{i=1}^{2p} {w_{ij}^{(h)} X_i } +b_j^{(h)}\right),
\end{equation}
where $g$ is the activation function for hidden layer neurons, and $w_{ij}^{(h)}$ connects the $i$-th input neuron and the $j$-th hidden neuron for
$i=1,2,...,2p$ and $j=1,2,...,J$. We use ${\rm {\bf w}}^{(o)}$ to denote the matrix of hidden layer weights $w_{ij}^{(h)}$,  and use ${\rm {\bf
b}}^{(h)}$ to denote the vector of corresponding bias terms $b_j^{(h)}$.

Unlike the CRM method which uses the center and range regression, the proposed method considers both the commonality and specialty of the relationship of the lower and upper bounds  inherently. That is, the hidden layer in this model extracts common features in lower and upper bounds, while the output layer captures the difference between these two bounds. Therefore, both individual pattern and cross-correlation between the upper and lower bounds
are taken into account by the proposed model.

Typically, the ANN models are optimized by minimizing the mean square error (MSE) loss function through gradient descent method. While for interval-valued
data prediction task, we have to also consider the mathematical coherence between the two predicted bounds such that  $\hat {Y}^U\ge
\hat {Y}^L$. In the extreme case, if the model fits data perfectly, then the coherence would definitely be satisfied. However, in practical
applications, the data may be complex and the coherence is not guaranteed. To meet this requirement, we add a non-crossing regularizer to the MSE
loss  and formulate the objective function as follows,
\begin{equation}
\label{eq12}
L=\frac{1}{2N}\sum\limits_{k=1}^N {(Y^L_k-\hat {Y}^L_k)^2}
+\frac{1}{2N}\sum\limits_{k=1}^N {(Y^U_k-\hat {Y}^U_k)^2} +\frac{\lambda
}{2N}\sum\limits_{k=1}^N {\left\{ {\max \left\{ {0,\hat {Y}^L_k-\hat {Y}^U_k} \right\}}
\right\}^2},
\end{equation}
where $\lambda >0$ is a regularization parameter. It is a combination of the prediction accuracy and the mathematical coherence between the two bounds. Clearly, the regularization term would be activated only when the estimated lower bound is greater than the corresponding upper bound, as $\hat {Y}^L-\hat {Y}^U>0$ would become a positive quantity. Through model training, the network would learn to minimize this additional penalty term to avoid the interval crossing phenomenon. On the contrary, if no interval crossing phenomenon is observed, the normal model training process would not be affected as the regularizer remains zero.

\subsection{Model Training}
In this paper, the backpropagation (BP) algorithm is employed to optimize the proposed RANN model. Given the loss function in (\ref{eq12}), the partial derivatives of
the loss function to the output layer weights and biases can be derived
according to the chain rule:
\begin{equation}\label{eq13}
\frac{\partial L}{\partial w_{j,L}^{(o)} }=\frac{\partial L}{\partial \hat
{Y}^L}\frac{\partial \hat {Y}^L}{\partial w_{j,L}^{(o)}}, \ 
\frac{\partial L}{\partial w_{j,U}^{(o)} }=\frac{\partial L}{\partial \hat
{Y}^U}\frac{\partial \hat {Y}^U}{\partial w_{j,U}^{(o)} }, \ 
 \mbox{ }\frac{\partial L}{\partial b^{(o)}_L }=\frac{\partial L}{\partial \hat
{Y}^L}\frac{\partial \hat {Y}^L}{\partial b^{(o)}_L }, \ 
\frac{\partial L}{\partial b^{(o)}_U }=\frac{\partial L}{\partial \hat {Y}^U}\frac{\partial \hat
{Y}^U}{\partial b^{(o)}_U }, 
\end{equation}
where the corresponding partial derivatives are 
\begin{eqnarray*}
\frac{\partial L}{\partial \hat {Y}^L} & = & \left[ {-(Y^L-\hat {Y}^L)+\lambda
\cdot \max \left\{ {0,\hat {Y}^L-\hat {Y}^U} \right\}} \right]\\
\frac{\partial L}{\partial \hat {Y}^U} & = &\left[ {-(Y^U-\hat {Y}^U)-\lambda
\cdot \max \left\{ {0,\hat {Y}^L-\hat {Y}^U} \right\}} \right].
\end{eqnarray*}
The other parts of derivatives can be evaluated by 
\begin{equation} \label{eq14}
\def\arraystretch{1.5}
\begin{array}{ll}
 \frac{\partial \hat {Y}^L}{\partial w_{j,L}^{(o)} }=f^{'}(({\bf w}^{(o)}_L)^T{\bf z} + b_L^{(o)})\cdot z_j , & 
 \frac{\partial \hat {Y}^L}{\partial b^{(o)}_L }=f^{'}(({\bf w}^{(o)}_L)^T{\bf z}  + b_L^{(o)}), \\
 \frac{\partial \hat {Y}^U}{\partial w_{j,U}^{(o)} }=f^{'}(({\bf w}^{(o)}_U)^T{\bf z}  +b_U^{(o)})\cdot z_j , & 
 \frac{\partial \hat {Y}^U}{\partial b^{(o)}_U }=f^{'}(({\bf w}^{(o)}_U)^T{\bf z} +b_U^{(o)}). 
 \end{array}
\end{equation}

Next, for the hidden layer weights and biases,  we can again apply the chain rule and derive the following formulas:
\begin{equation}
\label{eq15}
\frac{\partial L}{\partial w_{ij}^{(h)} }=\left[\frac{\partial L}{\partial \hat {Y}^L}\frac{\partial \hat {Y}^L}{\partial z_j }+\frac{\partial L}{\partial \hat {Y}^U}\frac{\partial \hat {Y}^U}{\partial z_j }\right]\frac{\partial z_j }{\partial w_{ij}^{(h)} }, \ 
 \frac{\partial L}{\partial b_j^{(h)} }=\left[\frac{\partial L}{\partial \hat {Y}^L}\frac{\partial \hat {Y}^L}{\partial z_j }+\frac{\partial L}{\partial \hat {Y}^U}\frac{\partial \hat {Y}^U}{\partial z_j }\right]\frac{\partial z_j }{\partial b_j^{(h)} }
\end{equation}
with the corresponding derivatives given by
\begin{eqnarray*}
\frac{\partial \hat {Y}^L}{\partial z_j} & = & f^{'}(({\bf w}^{(o)}_L)^T{\bf z} + b_L^{(o)})\cdot w_{j,L}^{(o)}\\
\frac{\partial \hat {Y}^U}{\partial z_j} & = & f^{'}(({\bf w}^{(o)}_U)^T{\bf z} + b_U^{(o)} )\cdot w_{j,U}^{(o)}\\
\frac{\partial z_j }{\partial w_{ij}^{(h)}}  & = & g^{'}(({\bf w}^{(h)}_j)^T{\bf X} +b_j^{(h)} ) \cdot X_i\\
\frac{\partial z_j }{\partial b_j^{(h)}}  & = & g^{'}(({\bf w}^{(h)}_j)^T{\bf X} +b_j^{(h)} )
\end{eqnarray*}

With the above derivative evaluations, the gradient descent algorithm could be easily utilized to update the model iteratively. Write ${\rm {\bf a}}\equiv ({\rm {\bf w}}_L^{(o)}, {\rm {\bf w}}_U^{(o)}, b_L^{(o)}, b_U^{(o)}, {\rm {\bf w}}^{(h)}, {\rm {\bf b}}^{(h)})$ that collects all the unknown parameters within the neural network, then the model could be optimized using the following iterative updating algorithm,
\begin{equation} \label{eq16}
{\rm {\bf a}}^{(t+1)}={\rm {\bf a}}^{(t)} - \rho_t\cdot {\rm {\bf \hat {s}}}^{(t)},
\end{equation}
where $\rho_t$ is the learning rate parameter and ${\rm {\bf \hat {s}}}^{(t)}$ is the vector of corresponding gradients at the step $t$. With proper choice of $\rho_t$, the algorithm is expected to reach a satisfying solution after sufficient iterations.  Here it is critical to select the adaptive learning rate parameter, for which we adopt the adaptive stochastic optimization algorithm ``Adam'' recently proposed by Kingma and Ba (2014).   
The Adam optimization method has been proved to be much more efficient and faster than othe  counterpart methods. 

The proposed RANN model is basically an interval-valued ANN model plus a
soft non-crossing regularizer, which fills in the research gap by
considering the interval's mathematical coherence property in machine
learning-based interval-valued prediction. As discussed in previous parts,
this model works for two reasons. First, using the powerful ANN structure,
this model is directly developed on the upper and lower bounds, where both
the individual pattern and cross-correlation are considered. Second,
compared with the inequality constraints used in other interval-valued
prediction methods, the proposed regularization method would not bring too
much harm to the prediction accuracy, as it only works when the actual
interval crossing occurs. In addition, by adjusting the regularization
parameter $\lambda $, both goals of the prediction accuracy and the mathematical coherence could be flexibly balanced. For example, for datasets
in which interval crossing phenomenon occurs frequently, we can choose a larger value of $\lambda $, and the model would place higher weights on preventing the interval crossing over MSE minimization, and vice visa.

\section{Experiments }
For illustration and verification purposes, we conduct experiments based on both simulated datasets and real-world datasets. In Subsections 4.1 and 4.2, we first descibe the experimental design with the evaluation metrics, benchmark models, and parameter settings. Then, two simulated datasets are introduced and tested in Subsection 4.3. Subsections 4.4 and 4.5 provide the experimental results of two real-world datasets. Subsection 4.6 gives the summary of the experimental results.

\subsection{Evaluation Metrics}
In order to evaluate the performance of different methods, four metrics are considered, including the root mean square error for lower bound ($RMSE_L$),
the root means square error for lower bound ($RMSE_U$), the mean Hausdorff
Distance ($MHD)$, and the coverage rate ($CR$).

The root mean square error (RMSE) is the most popular accuracy measurement in point prediction. For interval-valued data, we employ $RMSE_L $ and $RMSE_U $ to measure the fitting ability of the lower bound and upper bound, respectively, 
\begin{equation}
\label{eq15}
RMSE_L =\sqrt {\frac{\sum\limits_{k=1}^N {(Y_k^L -\hat {Y}_k^L )}^2 }{N}}, \mbox{ and }
RMSE_U =\sqrt {\frac{\sum\limits_{k=1}^N {(Y_k^U -\hat {Y}_k^U )}^2 }{N}}.
\end{equation}
The $MHD$ and $CR$ are responsible to evaluate the overall interval prediction ability. The difference lies in that $MHD$ is used to calculate
the distance between the predicted intervals and the true intervals,
\begin{equation} \label{eq16}
MHD=\frac{1}{N}\sum\limits_{k=1}^N {HD\left( {[Y_k^L ,Y_k^L ],[\hat {Y}_k^L, \hat {Y}_k^U ]} \right),}
\end{equation}
where the Hausdorff Distance is $HD\left( {I_1 ,I_2 } \right)=\max \left\{
{\mathop {\sup }\limits_{e_1 \in I_1 } \mathop {\inf }\limits_{e_2 \in I_2 }
\vert e_1 -e_2 \vert ,\mathop {\sup }\limits_{e_2 \in I_2 } \mathop {\inf
}\limits_{e_1 \in I_1 } \vert e_1 -e_2 \vert }\right\}$.
The $CR$ is used to calculate their overlap rates:
\begin{equation}
\label{eq17}
CR=\frac{1}{N}\sum\limits_{k=1}^N {\frac{\omega \left( {Y_k \cap \hat {Y}_k
} \right)}{\omega \left( {Y_k} \right)}}
\end{equation}
where $ \omega \left( {Y_k \cap \hat {Y}_k } \right) $ represents the overlap interval width of the predicted intervals and the true intervals.
Among the four performance criteria, smaller values are preferred for $MHD$, $RMSE_L$ and $RMSE_U$. In contrast, for the coverage rate $CR$, larger values would be better.

\subsection{Benchmark Models and Parameter Settings}
For comparison purpose, we include several benchmark models. For linear models, the classical CCRM method (Lima Neto and De Carvalho, 2010) and Lasso-IR model (Giordani, 2015) are considered. For nonparametric models, the IKRCR method (Fagundes, et al., 2014) is considered. For machine learning-based methods, the iMLP model (San Roque, et al., 2007) and MSVR model (P\mbox{\'{e}}rez-Cruz, et al., 2002; Xiong et al., 2014a) are considered.

For parameter settings, we follow the instructions provided in the literature or choose the best ones by cross-validation. Specifically, for the Lasso-IR model, we follow Giordani (2015) and choose the optimal
shrinkage control parameter $t$ via a 3-fold cross-validation based on the training data. For IKRCR, the bandwidth parameter of the Gaussian kernel is set to 0.1, as is suggested in the paper (Fagundes, et al., 2014). In terms of the
MSVR model, the most common RBF kernel is selected. Then, a 5-fold
cross-validation grid search method in the training data is employed to
select the best group of regularization parameter and kernel width, with the
grid $\{2^{-10},2^{-8},...,2^8,2^{10}\}\times
\{2^{-10},2^{-8},...,2^8,2^{10}\}$. 

For fair comparison, the two neural
network-based models, i.e., the iMLP model and the proposed RANN model, use
almost the same parameter settings. That is, both models use the identical
output activation function; the number of hidden neurons are empirically
determined within the range of 2 to 5, adjusted according to the datasets.
The popular stochastic optimization algorithm ``Adam'' algorithm (Kingma and Ba, 2014) is employed to optimize both models for
500 epochs, with the initial learning rates set to 0.001. The only
difference lies in the hidden layer activation, where the proposed RANN model uses the sigmoid function, while the iMLP model selects
"tanh" function as suggested by San Roque, et al. (2007). Finally, the additional regulization parameter $\lambda$ in the
proposed model is empirically set to 1 in all cases.

For ease of implementation, two programming languages are used
when conducting the experiments. The LassoIR model, the IKRCR
model, and the IKRCR model are implemented in Matlab, while the CCRM method is in
Python. The two neural work-based models are implemented via the powerful neural computing tool ``TensorFlow'' in Python.

\subsection{Simulation Studies}
We consider two data generation processes with different degrees of modeling difficulty: 
\begin{itemize}
\item Scenario 1: there is only one independent variable, and it is linearly related to the target variable; 
\item Scenario 2: there are two independent variables, and both have nonlinear relationships with the target variable.
\end{itemize}

\subsubsection{Scenario 1: Linear Model}
The first scenario uses a simple interval generation process, with one
independent variable and only linear relationship within data. A four-step
data generation procedure is involved.

\begin{enumerate}
\item Generate the center of independent variable $X^c \sim N(0,3)$;
\item Derive the center of target variable $Y^c =4+X^c +\varepsilon$, where $\varepsilon \sim N(0,1.5)$ denotes the white noise term;
\item Compute the half-range via $X^r =2-0.1X^c +\varepsilon _1 $ and $Y^r =1+0.1Y^c +\varepsilon _2 $, where $\varepsilon _1 \sim N(0,1.5)$ and $\varepsilon _2 \sim N(0,1.5)$ are white noise;
\item Finally, derive the lower and upper bounds of interval-valued data through the transformation: $X^L =X^c -X^r ,X^U =X^c +X^r ,Y^L =Y^c -Y^r ,Y^U =Y^c +Y^r $.
\end{enumerate}

\begin{figure}[htbp]
\centerline{\includegraphics[width=3in,height=2.8in]{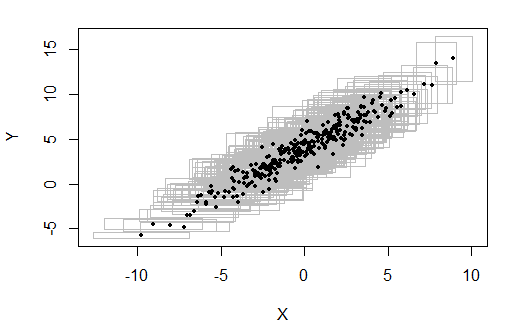}}
\caption{Interval-valued Data Denerated in Scenario 1}\label{fig2}
\end{figure}

With the above-mentioned method, we could obtain the first simulation data
by repeating step (\ref{eq1})--(\ref{eq4}) for 300 times. Thus, a data with 300 samples is
generated. For illustration purpose, one example of this data is shown in
Figure~\ref{fig2}.

\subsubsection{Scenario 2: Nonlinear Model}
To further test models' ability in handling nonlinear data, scenario two
considers two independent variables. Moreover, the centers and half-ranges
both follow certain nonlinear functions. The detailed data generation
procedure is as follows. 

\begin{enumerate}
\item Generate centers and half-ranges for the two independent variables respectively, i.e., $X_1^c \sim Unif(-1,1)$, $X_2^c \sim Unif(1,3)$, $X_1^r \sim Unif(0.5,1.0)$ and $X_1^r \sim Unif(1,1.5)$.
\item Derive the corresponding centers and half-ranges for target variable, with both quadratic and exponential relationship $Y^c =5\times e^{-(X_1^c )^2}+(X_2^c )^2+\varepsilon _1 $ and $Y^r =e^{-2\times (X_1^r )^2}+\frac{1}{2}\times (X_2^r )^2+\varepsilon _2 $, where $\varepsilon _1 \sim N(0,1)$ and $\varepsilon _2 \sim N(0,0.2)$.
\item Transform the center range values to interval-valued data $X^L =X^c -X^r ,X^U =X^c +X^r ,Y^L =Y^c -Y^r ,Y^U =Y^c +Y^r $.
\end{enumerate}

\begin{figure}[htbp]
\centerline{\includegraphics[width=6in,height=2.8in]{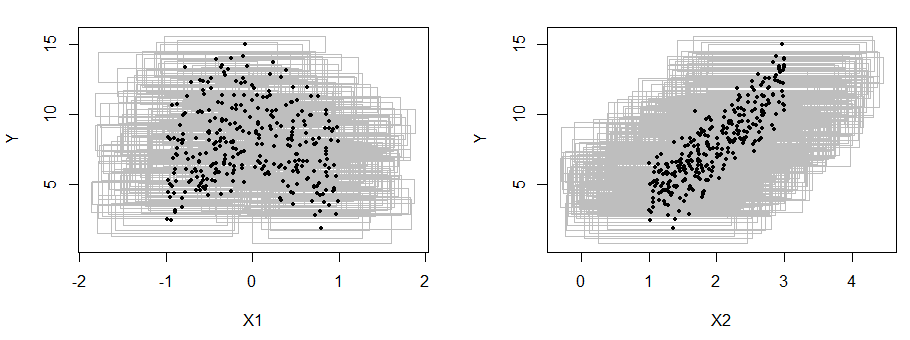}}
\caption{Interval-valued Data Generated in Scenario 2}
\label{fig3}
\end{figure}

We repeat these steps for 300 times and obtain the simulation data with 300 samples. One example is drawn in Figure~\ref{fig3}. Obviously,
this data is much more complicated than Scenario 1, with two mixed nonlinear patterns.

\subsubsection{Comparison Results}
Each simulation dataset is randomly split with 80{\%} for training and
20{\%} for testing. To alleviate randomness, the experiments are repeated
for 30 times. The averaged results are reported with the standard deviation shown in the brackets.

\begin{table}[htbp]
\caption{Experimental Results for Scenario 1}
\begin{center}
\begin{tabular}{c|ccc|c}
\hline
&
$MHD$&
$RMSE_L $&
$RMSE_U $&
$CR$ \\
\hline
CCRM&
1.054 (0.087)&
0.947 (0.093)&
1.143 (0.111)&
0.711 (0.028) \\
Lasso-IR&
2.201 (0.549)&
2.020 (0.549)&
2.196 (0.581)&
0.372 (0.125) \\
IKRCR&
1.078 (0.074)&
\textbf{0.903} (0.074)&
1.233 (0.096)&
0.700 (0.026) \\
MSVR&
\textbf{0.943} (0.100)&
0.930 (0.100)&
\textbf{1.136} (0.119)&
0.695 (0.037) \\
iMLP&
1.403 (0.090)&
1.004 (0.110)&
1.476 (0.113)&
\textbf{0.797} (0.033) \\
Proposed RANN&
0.955 (0.092)&
0.943 (0.092)&
1.151 (0.112)&
0.686 (0.035) \\
\hline
\end{tabular}
\label{tab1}
\end{center}
\end{table}

Table 1 lists the comparison results of the proposed RANN model with five benchmark models under Scenario 1. With respect to $MHD$, the MSVR model
performs the best followed by the proposed model. In terms of $RMSE_L$, $RMSE_U$ and $CR$, the three models of CCRM, IKRCR, MSVR and the proposed
model have very close results. The Lasso-IR model shows a significantly worse performance than all the other models, and the possible reason can be
referred to its use of inequality constraints, which leads to unsatisfying solutions.

\begin{table}[htbp]
\caption{Experimental Results for Scenario 2}
\begin{center}
\begin{tabular}{c|ccc|c}
\hline
&
$MHD$&
$RMSE_L $&
$RMSE_U $&
$CR$ \\
\hline
CCRM&
1.329 (0.095)&
1.400 (0.112)&
1.367 (0.095)&
0.514 (0.036) \\
Lasso-IR&
1.426 (0.180)&
1.576 (0.180)&
1.581 (0.217)&
0.482 (0.056) \\
IKRCR&
2.336 (0.172)&
2.521 (0.172)&
2.526 (0.165)&
0.286 (0.036) \\
MSVR&
1.162 (0.153)&
1.207 (0.153)&
1.192 (0.156)&
0.581 (0.060) \\
iMLP&
2.229 (0.159)&
2.010 (0.231)&
2.192 (0.213)&
0.598 (0.051) \\
Proposed RANN&
\textbf{1.073} (0.081)&
\textbf{1.118} (0.079)&
\textbf{1.104} (0.107)&
\textbf{0.602} (0.046) \\
\hline
\end{tabular}
\label{tab2}
\end{center}
\end{table}

Table 2 lists the comparison results under Scenario 2. The proposed RANN model achieves the best performance with respect to all the criteria, which demonstrates its effectiveness in handling nonlinear data. Following the proposed model, the MSVR model also performs well and beats the other models in most criteria. In contrast, even in nonlinear settings, the iMLP model and the IKRCR model have shown poor performances. We find that although the iMLP model uses the neural network structure, its overall performance is very weak, even worse than the linear CCRM model. The reason may be attributed to the use of absolute operation in iMLP neurons, which may lead to high biases.

\subsection{Mushroom Dataset}
\begin{figure}[htbp]
\centerline{\includegraphics[width=5in,height=2.8in]{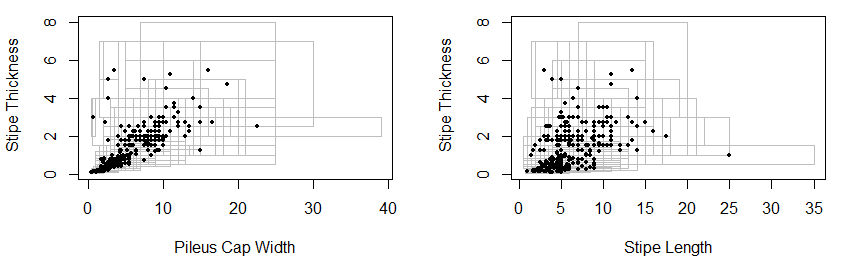}}
\caption{Interval-valued Plots of Mushroom Dataset}\label{fig4}
\end{figure}

The mushroom dataset is a famous interval-valued dataset which describes
different mushrooms species' appearance characteristics, including the
pileus cap width, the stipe length, and the stipe thickness. Typically, the
first two features are treated as independent variables, while the stipe
thickness is the dependent variable. All of these mushrooms belong to the
genus Agaricus and are extracted from the Fungi of California Species Index
(http://www.mykoweb.com/CAF/species\_index.html). 
We obtain the dataset from Xu (2010), and take 264 samples after omitting the missing values from the 274 observations. As shown in Figure~\ref{fig4},
these variables range from tiny mushrooms species to large ones, and the relationship among variables seems to be nonlinear.

\begin{table}[htbp]
\caption{Experimental Results on Mushroom Dataset}
\begin{center}
\begin{tabular}{c|ccc|c}
\hline
&
$MHD$&
$RMSE_L $&
$RMSE_U $&
$CR$ \\
\hline
CCRM&
0.747 (0.080)&
0.557 (0.115)&
1.085 (0.195)&
0.517 (0.036) \\
Lasso-IR&
0.742 (0.069)&
\textbf{0.538} (0.069)&
1.056 (0.170)&
0.456 (0.049) \\
IKRCR&
0.802 (0.069)&
0.543 (0.069)&
1.064 (0.184)&
0.489 (0.029) \\
MSVR&
1.018 (0.457)&
0.897 (0.457)&
1.516 (1.368)&
0.413 (0.112) \\
iMLP&
0.760 (0.112)&
0.578 (0.129)&
1.084 (0.241)&
\textbf{0.581} (0.080) \\
Proposed RANN&
\textbf{0.720} (0.099)&
0.545 (0.103)&
\textbf{1.015} (0.228)&
0.516 (0.062) \\
\hline
\end{tabular}
\label{tab3}
\end{center}
\end{table}

For model comparision, each model is run 30 times on the randomly split the data with 80{\%} for training and
20{\%} for testing. The numerical results in Table 3 show that the proposed RANN model performs the best in
terms of the minimum $MHD$ and $RMSE_U$. As for $RMSE_L$ and $CR$,  the proposed model also
ranks among the top three models. The two linear models of CCRM and Lasso-IR and the two nonlinear models of IKRCR and iMLP model achieve relatively
middle performance in most criteria, except for the iMLP obtains the largest $CR$ and the Lasso-IR gets the best $RMSE_L$. 
However, as a powerful machine learning-based model, the MSVR model shows the worst results, and a possible reason may be that the dataset has a large variable range and MSVR may fail to model such data structure.

\subsection{Hong Kong Air Quality Monitoring Dataset}
The Hong Kong air quality monitoring (HKAQM) dataset is released by Hong
Kong Environmental Department (http://epic.epd.gov.hk), which
aims at creating a healthy and clean environment for the next generation.
They provide hourly air quality data of 16 monitoring stations in Hong Kong.
We choose the data of Central Station and download the hourly data ranging
from Jan.~1, 2016 to Dec.~31, 2016. Then, we aggregate the hourly data to
the minimum and maximum form according to each day's record. This dataset
contains 7 variables. 
We choose the RSP (respirable suspended particulates) as the target variable while CO, NO2,
and SO2  as independent variables.  
Since HKAQM dataset is time-related, we may not randomly split the data. In such situation, we split
the data with the first 60{\%} as training set and the remaining 40{\%} as  testing set.
\begin{figure}[htbp]
\centerline{\includegraphics[width=6in,height=2in]{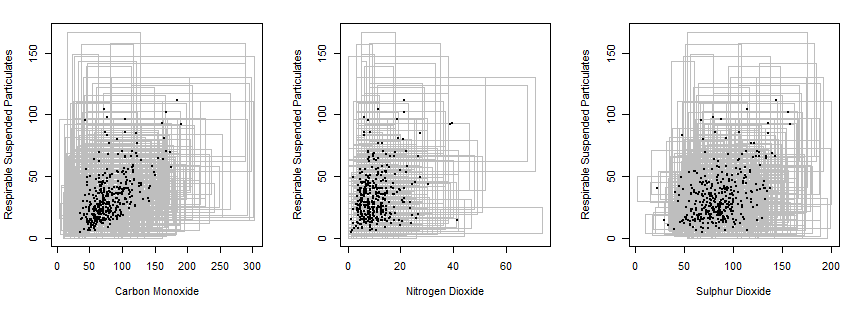}}
\caption{Interval-valued Plots of HKAQM Dataset}\label{fig5}
\end{figure}

\begin{table}[htbp]
\caption{Experimental Results on HKAQM Dataset}
\begin{center}
\begin{tabular}{c|ccc|c}
\hline
&
$MHD$&
$RMSE_L $&
$RMSE_U $&
$CR$ \\
\hline
CCRM&
18.883&
13.777&
19.888&
0.608 \\
Lasso-IR&
18.260&
13.541&
19.707&
0.636 \\
IKRCR&
17.781&
12.679&
19.471&
0.660 \\
MSVR&
20.634&
14.812&
21.804&
0.591 \\
iMLP&
16.941&
14.266&
\textbf{18.524}&
0.687 \\
Proposed RANN&
\textbf{16.840}&
\textbf{11.771}&
18.783&
\textbf{0.720} \\
\hline
\end{tabular}
\label{tab4}
\end{center}
\end{table}

In Table 4, we can see that the proposed model outperforms the other models
under the $MHD$, $RMSE_L$ and $CR$ criteria. As for $RMSE_U$, the proposed model is also very
close the best result  achieved by the iMLP model.
On the other hand, although the iMLP model wins in terms of $RMSE_U$ and performs well in
$MHD$ and $CR$, its performance in $RMSE_L$ is not that promising. The IKRCR model has an avarage performance which is superior to the linear models.
The MSVR model again ranks the last in this dataset.

\subsection{Results Summary}
According to the above experimental results, the following summaries can be made: 
\begin{enumerate}
\item For simple interval data with linear relationships, the proposed RANN model is at least comparable with its counterparts in most criteria; while for complicated interval data with nonlinear relationships, the proposed RANN model shows an overall improvement to the benchmark models;
\item The constraint-based methods, e.g., the Lasso-IR model and the iMLP model, show significant underperformance in certain datasets. It can be deduced that these constraints used for preventing interval crossing phenomenon may lead to the decrease of prediction accuracy. In contrast, with soft non-crossing regularization, no significant accuracy decrease is observed in our proposed RANN model.
\item Therefore, we can draw an important conclusion that the proposed RANN model can be used as a promising tool for interval-valued data prediction tasks, especially for complex datasets with  nonlinear relationships.
\end{enumerate}

\section{Conclusions}
This paper proposed a regularized artificial neural network (RANN) model for interval-valued data prediction. This model incorporates a non-crossing regularizer in the powerful neural network model, to reduce the interval crossing phenomenon. First, in terms of model fitting ability, the proposed model takes advantage of the powerful ANN structure and is able to handle complicated nonlinear problems. Second, unlike existing inequality constraint-based models such as CCRM and iMLP, the proposed regularization method is more flexible while retaining the prediction accuracy. Therefore, the proposed RANN model fills in the research gap between machine learning-based interval-valued prediction and the mathematical coherence of intervals.

Our experimental results on both simulation data and real-life data show that the proposed  RANN model is an effective tool in interval-valued prediction tasks, especially for complicated nonlinear datasets. Our RANN model shows better performance than its counterparts, such as the iMLP model and the MSVR model in most cases. It also shows its superiority over the linear models when the data is complex and nonlinear.  However, as the neural network methods usually cost more time for model training than linear models, it is suggested to use the simple linear models for simple dataset with potentially linear relatinships. 

Our future research would be focused on the selection of the regularization parameter in the RANN model, in particular the data-driven approach like the cross validation. Also, with the fast development of deep learning techniques, it becomes possible for us to extend the RANN model to deep neural networks with multiple layers. Finally, we are also interested in developing the neural network-based interval prediction models for some more challenging tasks, e.g. financial interval time series forecasting.

\section{Acknowledgements}

This research project was  partially supported by Basic Research Seed Fund (201611159250) and Big Data Project Fund of The University of Hong Kong.

\section{References}
\begin{enumerate}
\item Billard, L., Diday, E., 2000. Regression analysis for interval-valued data. In~\textit{Data Analysis, Classification, and Related Methods}~(pp. 369-374). Springer, Berlin, Heidelberg.
\item Billard, L., Diday, E., 2002. Symbolic regression analysis.~\textit{Classification, Clustering, and Data Analysis} 281-288.
\item Fagundes, R.A., De Souza, R.M., Cysneiros, F.J.A., 2014. Interval kernel regression.~\textit{Neurocomputing}~$128$, 371-388.
\item Giordani, P., 2015. Lasso-constrained regression analysis for interval-valued data.~\textit{Advances in Data Analysis and Classification}~$9$(\ref{eq1}), 5-19.
\item Hao, P., Guo, J., 2017. Constrained center and range joint model for interval-valued symbolic data regression.~\textit{Computational Statistics {\&} Data Analysis}~116, 106-138.
\item Jeon, Y., Ahn, J., Park, C., 2015. A nonparametric kernel approach to interval-valued data analysis.~\textit{Technometrics}~$57$(\ref{eq4}), 566-575.
\item Kingma, D., Ba, J., 2014. Adam: A method for stochastic optimization.~arXiv preprint arXiv:1412.6980.
\item Lima Neto, E.A., De Carvalho, F.A.T., 2008. Centre and range method for fitting a linear regression model to symbolic interval data.~\textit{Computational Statistics {\&} Data Analysis}~$52$(\ref{eq3}), 1500-1515.
\item Lima Neto, E.A., De Carvalho, F.A.T., 2010. Constrained linear regression models for symbolic interval-valued variables.~\textit{Computational Statistics {\&} Data Analysis}~$54$(\ref{eq2}), 333-347.
\item Lim, C., 2016. Interval-valued data regression using nonparametric additive models.~\textit{Journal of the Korean Statistical Society}~$45$(\ref{eq3}), 358-370.
\item Maia, A.L.S., de Carvalho, F.D.A., 2011. Holt's exponential smoothing and neural network models for forecasting interval-valued time series.~\textit{International Journal of Forecasting}~$27$(\ref{eq3}), 740-759.
\item Maia, A.L.S., de Carvalho, F.D.A., Ludermir, T.B., 2008. Forecasting models for interval-valued time series.~\textit{Neurocomputing}~$71$(\ref{eq16}), 3344-3352.
\item P\'{e}rez-Cruz, F., Camps-Valls, G., Soria-Olivas, E., P\'{e}rez-Ruixo, J.J., Figueiras-Vidal, A.R. and Art\'{e}s-Rodr\'{\i}guez, A., 2002, August. Multi-dimensional function approximation and regression estimation. In~\textit{International Conference on Artificial Neural Networks}~(pp. 757-762). Springer, Berlin, Heidelberg.
\item San Roque, A.M., Mat\'{e}, C., Arroyo, J., Sarabia, \'{A}., 2007. iMLP: Applying multi-layer perceptrons to interval-valued data.~\textit{Neural Processing Letters}~$25$(\ref{eq2}), 157-169.
\item Xiong, T., Bao, Y., Hu, Z., 2014a. Multiple-output support vector regression with a firefly algorithm for interval-valued stock price index forecasting.~\textit{Knowledge-Based Systems}~$55$, 87-100.
\item Xiong, T., Bao, Y., Hu, Z., 2014b. Interval forecasting of electricity demand: a novel bivariate EMD-based support vector regression modeling framework.~\textit{International Journal of Electrical Power {\&} Energy Systems}~$63$, 353-362.
\item Xu, W., 2010.~\textit{Symbolic data analysis: interval-valued data regression}~(Doctoral dissertation, University of Georgia).
\item Xu, S., An, X., Qiao, X., Zhu, L., Li, L., 2013. Multi-output least-squares support vector regression machines.~\textit{Pattern Recognition Letters}~$34$, 1078-1084.
\end{enumerate}

\end{document}